\documentstyle[11pt,newpasp,twoside,graphicx,amstex]{article}
\def\deg{^\circ}

\def\procevn6{in Proc.\ of the 6th EVN Symp., ed.\ E.\ Ros
et al., (Bonn: MPIfR)}

\def\edcomment#1{\iffalse\marginpar{\raggedright\sl#1\/}\else\relax\fi}
\reversemarginpar

\begin{document}
\title{High Precision Differential Astrometry}
\author{Eduardo Ros}
\affil{Max-Planck-Institut f\"ur Radioastronomie, 
Bonn, Germany}

\begin{abstract}
While conventional imaging in VLBI
provides information only about the
relative position between different features of a given source,
phase-referenced observations can provide precise positional information
with respect to an external reference. The use of differential techniques
relies on the fact that most of the instrumental and propagation effects
are common for radio sources sufficiently close together in the sky. 
These effects cancel by subtracting the visibility phases
of the reference source from that of the neighboring
objects. Differential VLBI astrometry yields relative positions of radio
sources with sub-milliarcsecond accuracy. 
These high precisions can be applied to solve
various astrophysical problems. 
In this paper I will review
the progress made recently in this field, both in technique and science.
\end{abstract}

\section{The technique}

Astrometry is the domain of astronomy
devoted to the determination of positions and their time-variations.
In the radio domain, sub-milliarcsecond accuracy
is reached through high-precision very-long-baseline
interferometry (VLBI), using the positional information
obtained from the fringe phases.

The theoretical positional precision of an interferometer is 
$\sigma_{\rm R.A.,\,Dec} = (1/2\pi)\times(1/{\rm SNR})\times(\lambda/D)$,
where SNR is the signal-to-noise ratio, $\lambda$ is the wavelength, 
and $D$ is the baseline length (Lestrade et al.\ 1990).  
For SNR$\sim$15, $D=8000$\,km, and $\lambda$=3.6\,cm,
a precision of 10\,$\mu$as is possible;
$\lambda$\,1.3\,cm yields
4\,$\mu$as, and $\lambda$\,0.7\,cm, 2\,$\mu$as. 
Comparable precisions will be obtained by the optical, astrometric
satellites, such as the {\sl Space Interferometric Mission} ({\sl SIM}), 
with 4\,$\mu$as in
pointed mode, or {\sl GAIA}, with 1\,(10)\,$\mu$as for magnitudes of
5\,(10) in survey mode.
The Atacama Large Millimetre Array (ALMA) will 
have $D=10$\,km, and for a SNR$\sim$15 and 
$\lambda$\,0.87\,mm, a precision of 200\,$\mu$as
will be obtained, much poorer than VLBI values.

The main astrometric observable is the delay, directly
related to the fringe phase as a function of frequency.
It can be described as 
$\tau_{\rm geom}=\vec{D}\cdot \hat{s}/c$, where
$\vec{D}$ is the baseline vector and $\hat{s}$ a unit
vector towards the observed source.  It is directly related
to the total phase $\phi_{\rm T}$ from the 
interferometer output 
(visibility function with amplitude and
phase, $A\cos\phi_{\rm T}$, $\phi_{\rm T}\sim2\pi\nu \tau$).

To determine relative positions in the sky, two 
neighboring sources are observed and their relative phases
are determined, either simultaneously 
if they are both within
the primary antenna beam of each element
of the interferometer, or alternately
if switching from one source to the other is needed.
The observation of neighboring sources cancels out most of the systematic
effects in the astrometric data reduction, since they are mostly common
for both.
We analyze the observable of one source subtracted from the other,
that is, we perform {\sl differential} astrometry.
Most of the geometrical parameters involved in the observation
are taken account of and removed during the correlation process,
leaving one to work with the residuals.


The relevant residual quantities measured are
the phase delay as $\tau_{\phi}=
(\phi_{\rm T} + 2\pi n)/(2\pi \nu)$, where $n$ is an unknown integer;
the group delay which gives the change 
of phase with frequency,
$\tau_{\rm G}=(1/2\pi)\times(\partial \phi_{\rm T}/\partial \nu)$;
and
the delay rate, which measures the change of delay with time,
as $\dot{\tau}=(1/2\pi\nu)\times(\partial \phi_{\rm T}/\partial t)$.
The group delay and the delay rate are unambiguous observables, but
less precise than the phase-delay, which is $n/\nu$-ambiguous.


\subsection{Phase-reference mapping vs.\ phase-delay astrometry}

The residual, differential phase can be split in the following
terms:
\[
\begin{array}{ccccccccc}
\phi_{{\rm A}-{\rm B}}^{\rm res}&=& 
\phi_{{\rm A}-{\rm B}}^{\rm res,\,\sl str} &+&
\phi_{{\rm A}-{\rm B}}^{\rm res,\,\sl pos} &+&
\phi_{{\rm A}-{\rm B}}^{\rm res,\,\sl inst} &+&
\phi_{{\rm A}-{\rm B}}^{\rm res,\,\sl prop} \,,\\[4pt]
 && \scriptsize{(1)} && \scriptsize{(2)} && \scriptsize{(3)} && \scriptsize{(4)} \\
\end{array}
\]
namely, the {\sl str}uctural term (caused by the fact that radio sources
are not point-like), the {\sl pos}itional term, 
the {\sl inst}rumental 
term (antenna
electronics, clocks, etc.), and the unmodeled term of the 
{\sl prop}agation medium
(atmosphere, ionosphere).  

\paragraph{Phase-delay astrometry.}
Assuming both sources are strong enough to be detected,
if we provide an {\sl a priori} model for (3) \& (4), and we remove
(1) by using hybrid maps from the two sources, we determine (2) (and
corrections for (3) and (4)) via
a weighted least-squares fit.  This process also needs 
to solve for the phase ambiguities, which is known as
{\sl phase-connection} (Shapiro et al.\ 1979).  
Historically, this was the first
approach to perform high precision differential astrometry.

\paragraph{Phase-reference mapping.}
Assuming that the residuals 
(3) and (4) are sufficiently small, 
the phase from
one source (the strong one) is interpolated
to the other one and a Fourier-transform is performed:
\begin{equation*}
|V(u,v)|e^{i\phi_{{\rm A}-{\rm B}}} \longrightarrow
I(x,y)=\int\int V(u,v)e^{-i2\pi(ux+vy)}\,du\,dv\,,
\end{equation*}
where $(x,y)$ are the coordinates in the image plane, and
$(u,v)$ their Fourier-pairs, $V$ is the visibility function,
and $I$ is the brightness distribution.
By this process an image with some offset position from the origin
is obtained, recovering
also the structure of the source.  
The first phase-reference map with in-beam observations
(1038+528\,A/B, 33$^{\prime\prime}$ separation)
was published by Marcaide \& Shapiro (1984).  The first
one from switched observations
(0249+436/0248+483, 0$\rlap{.}\deg$5 sep.)
was presented in Alef (1988).  

A method related to the phase-reference mapping 
is the hybrid double mapping (see Rioja \& Porcas 2000), 
where
the visibility functions of two nearby sources are added together
and both are imaged in the same field; therefore, the offset in one
of the images w.r.t.\ the other in the map is the difference between
their {\sl a priori} coordinate differences and the `true' value.
The fast-frequency switching presented by Middelberg et al.\ (2002)
interpolates the phase from one frequency to another frequency by 
multiplying by the frequency ratio.

The cluster-cluster (or multi-view) method, first presented by Counselman et al.\
(1974), uses a common local oscillator for multiple antenna elements 
(a ``cluster") in different sites.  This enables simultaneous observations
of multiple sources to be made
on the separate ``sub-baselines" between sub-elements from
one site to another.  Recently, Rioja et al.\ (2002) reported on successful
observations at 1.6\,GHz.

\subsection{Recent technical achievements}

In recent years, some progress has allowed highest
astrometric precisions to be obtained, reaching higher frequencies, etc.  
One key aspect is the removal of the delay term introduced by
the ionospheric plasma, which is frequency-dependent.  Classical
geodetic experiments use dual-band observations (2.3/8.4\,GHz) 
to account for this.  The development of the Global Positioning System (GPS)
in the 1990s made it possible
to estimate the total 
electron content (TEC) between a GPS receiver site and a 
dual-frequency transmitting satellite.
Using measurements made
at sites near to VLBI sites, it is possible to estimate the ionospheric
contribution to the VLBI observables.  This was shown by Ros et al.\ (2000),
and now the ionospheric correction can be performed routinely from 
global IONEX TEC data within
{$\cal A$$\cal I$$\cal P$$\cal S$} using the task {\tt tecor} (see, e.g., Walker 
\& Chatterjee 1999).

In phase-delay astrometry, phase-connection has been extended up
to distances of 15$\deg$ (P\'erez-Torres et al.\ 2000), and up to frequencies
of 43\,GHz (Guirado et al.\ 2000).  A phase-referencing test at 86\,GHz 
has also been successful (Porcas \& Rioja 2002).
Fomalont \& Kopeikin (2002) and Fomalont et al.\ (these proceedings)
have shown that, using multiple calibrators at 8.4\,GHz, precisions below
10\,$\mu$as can be obtained by phase-referencing.

The application of astrometry to orbiting VLBI has been limited by the
fact that the satellite {\sl HALCA} could not observe in nodding mode.
In-beam observations at 5\,GHz have been successful (Porcas et al.\ 2000; 
Guirado et al.\ 2001), yielding an upper bound of 10\,m to the
uncertainty of the spacecraft orbit reconstruction.

\section{The Science}

The relative angular positions of extragalactic radio sources inferred
from VLBI astrometry form the best realization of an inertial reference
frame in astronomy.  A continuous program of monitoring hundreds of 
radio sources establishes and maintains two
reference frames: the terrestrial (antenna positions with mm-accuracy) 
and the celestial (see, e.g., Sovers et al.\ 1998).  
The International Astronomical Union
adopted the International Celestial Reference Frame (ICRF) as the fundamental
one (see Ma et al.\ 1997, 1998).  The ICRF
improves the precision of the latest optical ``fundamental catalogue", the
FK5 (Fricke et al.\ 1988) by more than one order of magnitude.  Special
techniques must be used to connect the VLBI celestial frame to the
historical optical frame (Lestrade et al.\ 1995).  This is of special
importance with the optical satellites like {\sl HIPPARCOS}, {\sl SIM},
or {\sl GAIA}, which have a precision equal to, or better
than VLBI.

The accurate determination of positions of celestial bodies has
many applications in radio astronomy.  
Astrometry of stellar masers is covered
in the review given by Boboltz (these proceedings).
High-precision astrometry has been applied to determine
the positions, proper motions, and parallaxes of pulsars (see, e.g., 
Campbell et al.\ 1996 and Brisken, these proceedings).  
The registration of the young
supernova remnant SN\,1993J
w.r.t.\ the nucleus of M\,81 (Bietenholz et al.\ 2001) gave the position of the
dynamical center of the shell emission and limited any possible one-sided
expansion of the shell to 5.5\%.
VLBI astrometry has also been applied to determine the Galactic Center 
position w.r.t.\ J1745$-$285 and J1748$-$291 at 43\,GHz, yielding
an estimate of its proper motion 
and a lower limit to its
mass (see Reid et al., these proceedings).  

The astrometric technique also allows some
of the predictions of general relativity to be tested, 
by analyzing the effect on the incoming signals
by their passage through the varying gravitational potential
within the Solar System.  
Lebach et al.\ (1995) determined the value
$\gamma_{\rm PPN}=0.9996\pm0.0017$ for the parameterized
post-Newtonian relativity theory, from group-delay observations of
3C\,279 w.r.t.\ 3C\,273 during a solar occultation.  
Recently, the speed of gravity\footnote{There is a controversy whereby
some authors claim this to be just the speed of light.} 
has been measured during the September
2002 Jupiter conjunction with the QSO J0832+1835, providing the parameter
$\delta=(c/c_{\rm grav})-1=-0.02\pm0.19$ 
that yields a value of
$c_{\rm grav}=1.06\pm0.21c$ (Fomalont \& Kopeikin 2003).
The detection of the frame dragging 
of the terrestrial gravitational potential 
(expected to be 42\,mas\,yr$^{-1}$)
is the purpose of the
{\sl Gravity Probe B} mission, where IM\,Pegasi (HR\,8703) will
be used as a guide star.  Astrometric observations have been carried
out on this flaring star since 1997 to fix its astrometric parameters
within 150\,$\mu$as\,yr$^{-1}$
(see Lebach et al.\ 1999 and Ransom et al., these
proceedings).
Honma \& Kurayama (2002)
suggest that VLBI astrometry could even be used to study 
gravitational micro-lensing.

\paragraph{AGN studies.}
Following the standard jet model in active galactic
nuclei (AGN), the opacity $\sim1$ surface (a.k.a.\ the
{\sl core}) has a frequency-dependent position.  
The registration of close pairs
of AGN allows the stability of the core in time and 
its changes with frequency to be determined.
For instance, Bartel et al.\ (1986) showed that the core in 3C\,345 is stable
within 20\,$\mu$as\,yr$^{-1}$.  Marcaide \& Shapiro (1984) proved
that the core position
in the QSO 1038+528\,A is frequency-dependent, by determining
its position w.r.t.\ 1038+528\,B via in-beam observations.  
The upper limit on the proper motion of one core w.r.t.\ to the other
is 10\,$\mu$as\,yr$^{-1}$ (Rioja \& Porcas 2000).
Guirado et al.\ (1995) determined the position of 4C\,39.25 
w.r.t.\ the reference source 0920+390 
and could establish a proper motion of its B component of 
$90\pm43$\,$\mu$as\,yr$^{-1}$ in R.A.\ and 
$7\pm68$\,$\mu$as\,yr$^{-1}$ in Dec., 
confirming the nature of this component as a shock wave.
This result was later confirmed by the geodetic data presented 
in Fey et al.\ (1997).
Ros et al.\ (1999) compared the estimated position of 1928+738 w.r.t.\ 
2007+777 from epochs 1985 to 1992, and observed changes in the position of
the VLBI core.  This implies that the dynamical center of this QSO is 
to the north of the core.  

These observations, together with the extension
of the phase connection technique to distances of 15$\deg$
(P\'erez-Torres et al.\ 2001), has lead to a 
project of observations to
establish the absolute kinematics of radio source components in the
thirteen radio sources of the S5 polar cap sample
(first images are presented in Ros et al.\ 2001), which is being carried
out at frequencies of 8.4, 15, and 43\,GHz.  These phase-delay
observations use bootstrapping techniques and the geometrical constraints
given by the location of all the radio sources within 30$\deg$ of 
the north celestial pole.

\paragraph{Stellar astrometry --- search for exo-planets.}
We have mentioned above the importance of the link between the 
reference frame provided by {\sl HIPPARCOS} and VLBI measurements.
This has been done by Lestrade et al.\ (1995, 1999) using 11 radio stars
in the northern and southern hemispheres.  One of the stars initially chosen
for this work was the flaring AB\,Doradus, 
which exhibited large
residuals after solving for proper motion and parallax.  Guirado
et al.\ (1997) presented a detailed
analysis combining {\sl HIPPARCOS} and VLBI solutions, showing the presence
of a low mass companion of 0.08$-$0.11\,M$_\odot$ orbiting the
catalogue star (of 0.76\,M$_\odot$).  This experiment demonstrates the
feasibility of detecting brown dwarfs and exo-planets orbiting
stars at tens of parsecs.

The radial velocity method (based on Doppler shift measurements) 
of detecting exo-planets favors the detection of
objects at short orbital radii and large speeds.  In contrast, 
astrometric measurements of the ``wobble" favor the detection of planets at
large orbital radii (which need longer observing campaigns, since the
orbital periods can be tens of years).  In Fig.\ 1 we show in detail the 
region accesible to VLBI astrometry.  A program of this kind is being
carried out using the small but sensitive array consisting of
the DSN antennas and Effelsberg, observing nearby M dwarfs with a resolution
of 1\,mas (first results are
shown in Guirado et al.\ 2002).

\begin{figure}[t!]
\vspace{-10pt}
\begin{center}
\includegraphics[clip,width=0.84\textwidth]{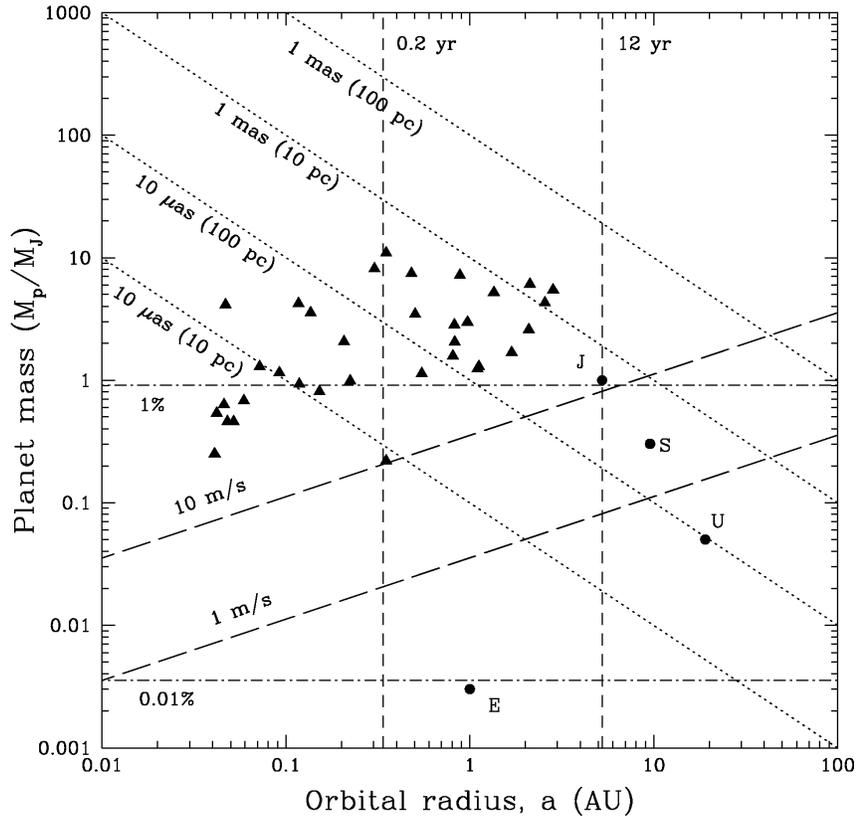}
\end{center}
\vspace{-10pt}
\caption{
\small
Detection domains for methods aimed at detecting exo-planet orbital motion,
assuming
M$_\star$\,=\,M$_\odot$.
Triangles design
lower limits on the masses of known planetary systems, and
circles denote 
Jupiter (J), Saturn (S), Uranus (U), and the Earth (E).
The diagonal lines show the exo-planet detection limits
for radial velocity and wobble measurements.
Horizontal lines show photometric
detection thresholds for planetary transits
of 1\% and 0.01\%, corresponding roughly to Jupiter and the Earth.
The region below 10\,m\,s$^{-1}$ and above
10\,$\mu$as (at 10\,pc) is `reserved'
for exploration in astrometry.
{\sl Figure taken from Perryman, 
Rep.\ Prog.\ Phys., 63, (2000).}
}
\end{figure}

\section{The Future}

The construction of the VLBA facilitated 
astrometric observations in the last decade, thanks to
the high slewing speeds of the antennas for nodding between sources,
and the automization of the data reduction.  
Together with the advent of the {\sl SIM} and {\sl GAIA} era, high
resolution radio astrometry will also experience progress during the
coming years.
Improvement of the geometrical
models will play a role (better determination of polar motion,
the mapping functions which describe the behavior of the
atmospheric delay as a function of elevation, antenna positions, etc.).
Progress is being made in modeling the propagation medium,
using water vapor radiometers and improved GPS analyses.
A more timely correlation (e.g., by eVLBI, see Garrington, these 
proceedings) will facilitate more rapid results for of the length of
the day and other geodetic/astrometric parameters.
The implementation of nodding capabilities in future space VLBI missions
would also enhance the astrometric precision for sources separated by
a few degrees, given the higher resolution provided by baselines longer
than the Earth size (assuming that the spacecraft orbit parameters
are well-determined).
The continuous
improvement in antenna performance and new elements being added to
the VLBI networks will provide more and better data.  Finally, a
new arrays designed for astrometry is underway, the VERA project
(Kobayashi, these proceedings).
For the planned Square Kilometer Array (SKA) telescope,
a set of interesting suggestions for high-resolution options for astrometry 
has been proposed by Charlot (2001).

\small
\acknowledgments
I am grateful to R.\ W.\ Porcas, J.\ C.\ Guirado, W.\ Alef, and M.\ Kadler for
their careful reading and helpful comments on this manuscript.


\begin{references}

\reference Alef, W.
1988, in IAU Symp. 129, The Impact of VLBI on Astrophysics and Geophysics, 
ed.\ M.\ J.\ Reid \& J.\ M.\ Moran (Dordrecht: Kluwer), 523 


\reference Bartel, N., et al.
1986, Nature, 319, 733 



\reference Bietenholz, M.\ F., Bartel, N., \& Rupen, M.\ P.
2001, \apj, 557, 770 




\reference Campbell, R.\ M., et al.
1996, \apj, 461, L95

\reference Charlot, P.
2001, Astrometry with the SKA, in High Resolution Options for the SKA,
Bonn, 10-11 December 2001,
{\scriptsize \verb"http://www.euska.org/workshops/hr_ws_MPIfR_Bonn.html"}


\reference Counselman, C.\ C., et al.
1974,
\prl, 33, 1621


\reference Fey, A.\ L., Eubanks, M., \& Kingham, K.\ A.
1997, \aj, 114, 2284

\reference Fricke, W., et al.
1988, Fifth fundamental catalogue (FK5). Part 1: The basic fundamental stars,
(Heidelberg: Astronomische Rechen-Inst.)


\reference Fomalont, E.\ B., \& Kopeikin, S.
2002,
\procevn6, 53

\reference Fomalont, E.\ B., \& Kopeikin, S.
2003,
astro-ph/0302294



\reference Guirado, J.\ C., et al.
1995, \aj, 110, 2586

\reference Guirado, J.\ C., et al.
1997, \apj, 490, 835


\reference Guirado, J.\ C., et al.
2000, \aap, 353, L37 

\reference Guirado, J.\ C., et al.
2001, \aap, 371, 766

\reference Guirado, J.\ C., et al.
2002, \procevn6, 255

\reference Honma, M., \& Kurayama, T.
2002, \apj, 568, 717 



\reference Lebach, D.\ E., et al.
1995, \prl, 75(8), 1439

\reference Lebach, D.\ E., et al.
1999, \apj, 517, L43

\reference Lestrade, J.-F., et al.
1990, \aj, 99, 1663


\reference Lestrade, J.-F., et al.
1995, \aap, 304, 182

\reference Lestrade, J.-F., et al.
1999, \aap, 344, 1014

\reference Ma, C., et al.
1997, in IERS Technical Note 23, ed. C. Ma \& M. Feissel (Paris: Observatoire
de Paris)

\reference Ma, C., et al.
1998, \aj, 116, 516


\reference Marcaide, J.\ M., \& Shapiro, I.\ I.
1984, \apj, 276, 56



\reference Middelberg, E., et al.
2002, \procevn6, 61


\reference P\'erez-Torres, M.\ A., et al.
2000, \aap, 360, 161

\reference Perryman, M.\ A.\ C.
2000, Rep.\ Prog.\ Phys., 63, 1209 

\reference Porcas, R.\ W., et al.
2000,
in Astrophysical Phenomena Revealed by Space VLBI, ed.\ H.\ Hirabayashi,
P.\ G.\ Edwards, \& D.\ W.\ Murphy, (Tokyo: ISAS), 245 

\reference Porcas, R.\ W., \& Rioja, M.\ J.
2002, \procevn6, 65




\reference Rioja, M.\ J., \& Porcas, R. W.
2000, \aap, 355, 552


\reference Rioja, M.\ J., et al.
2002, \procevn6, 57


\reference Ros, E., et al.
1999, \aap, 348, 381 

\reference Ros, E., et al.
2000, \aap, 356, 357

\reference Ros, E., et al.
2001, \aap, 376, 1090

\reference Shapiro, I.\ I., et al.
1979, \aj, 84, 1459 


\reference Sovers, O.\ J., Fanselow, J.\ L., \& Jacobs, C.\ S.
1998, Rev.\ Mod.\ Phys., 70, 1393 

\reference Walker, C., \& Chaterjee, S.,
1999,
Ionospheric Corrections using GPS Based Models,
VLBA Scientific Memo 23, (Socorro, NM: NRAO)

%
\end{references}
\end{document}